\documentclass[apl, superscriptaddress, preprint]{revtex4-1}
\usepackage{graphicx}% Include figure files
\usepackage{bm}% bold math

\usepackage{hyperref}% add hypertext capabilities
\hypersetup{
            colorlinks=true,
            linkcolor=blue,
            anchorcolor=blue,
            citecolor=blue
            }
\usepackage{changes}
\usepackage{amsmath, amssymb}

\begin{document}

%\preprint{APS/123-QED}
%\title{Normal/inverse Doppler effect of backward volume magnetostatic spin waves } 
\title{Distinguishing Backward Volume Magnetostatic Spin Wave Vectors via the Spin Wave Doppler Effect } 
%\author{Shaojie Hu}
%\email[]{shaojiehu@mail.xjtu.edu.cn}
%\affiliation{Center for Spintronics and Quantum Systems, State Key Laboratory for Mechanical Behavior of Materials, School of Materials Science and Engineering, Xi'an Jiaotong University, Xi'an, Shaanxi, 710049, China}

\author{Xuhui Su}
%\email[]{sxh813@stu.xjtu.edu.cn} 
\affiliation{College of Integrated Circuits and Optoelectronic Chips, Shenzhen Technology University, 3002 Lantian Road, Pingshan District, Shenzhen Guangdong, China, 518118}
\affiliation{School of Microelectronics $\&$ State Key Laboratory for Mechanical
Behavior of Materials, Xi'an Jiaotong University, Xi'an 710049, China }
%\affiliation{Center for Spintronics and Quantum Systems, State Key Laboratory for Mechanical Behavior of Materials, School of Materials Science and Engineering, Xi'an Jiaotong University, Xi'an, Shaanxi, 710049, China}

\author{Dawei Wang}
%\email[]{dawei.wang@xjtu.edu.cn}
\affiliation{School of Microelectronics $\&$ State Key Laboratory for Mechanical Behavior of Materials, Xi'an Jiaotong University, Xi'an 710049, China }

\author{Shaojie Hu}
%\email%[Authors to whom correspondence should be addressed:\ ]{hushaojie@sztu.edu.cn}
%\email[]{hu.shaojie@phys.kyushu-u.ac.jp} 
%\affiliation{Department of Physics, Kyushu University, 744 Motooka, Fukuoka, 819-0395, Japan}
\email[]{hushaojie@sztu.edu.cn} 
\affiliation{College of Integrated Circuits and Optoelectronic Chips, Shenzhen Technology University, 3002 Lantian Road, Pingshan District, Shenzhen Guangdong, China, 518118}

\date{\today}

\begin{abstract}
Spin waves (SWs) and their quanta, magnons, are essential to achieving low-power information transmission in future spintronic devices. Backward volume magnetostatic spin waves (BVMSWs) exhibit a unique dispersion relationship: one frequency corresponding to two distinct wave vectors. At low wave numbers, dipole-dipole interactions dominate, resulting in negative group velocities, whereas at high wave numbers, exchange interactions prevail, producing positive group velocities. This dual behavior complicates wave vector identification and obscures intrinsic spin-wave interactions. In this study, we propose an approach based on the spin wave Doppler effect to effectively distinguish different wave vectors. At low wave numbers, the inverse Doppler effect occurs due to antiparallel phase and group velocities, while at high wave numbers, a normal Doppler effect emerges from parallel velocities. This method not only clarifies the underlying spin-wave interactions but also  help mitigate serious interference issues in the design of spin logic circuits.

%Spin waves (SWs) and their quanta, magnons, play a crucial role in enabling low-power information transfer in future spintronic devices. In backward volume magnetostatic spin waves (BVMSWs), the dispersion relation shows a negative group velocity at low wave numbers due to dipole-dipole interactions and a positive group velocity at high wave numbers, driven by exchange interactions. This duality complicates the analysis of intrinsic interactions by obscuring the clear identification of wave vectors. Here, we offer an innovative approach to distinguish between spin waves with varying wave vectors more effectively by spin wave Doppler effect. The spin waves at low wave numbers display an inverse Doppler effect because their phase and group velocities are anti-parallel. Conversely, at high wave numbers, a normal Doppler effect occurs due to the parallel alignment of phase and group velocities. Analyzing the spin wave Doppler effect is essential for understanding intrinsic interactions and can also help mitigate serious interference issues in the design of spin logic circuits.

\end{abstract}
\maketitle
Spin waves (SWs) are collective excitations in magnetic materials, propagating through exchange or dipolar interactions among precessing spins \cite{Bloch1932zur, Holstein1940field, Dyson1956general}. Recently, SWs have attracted significant attention for their potential applications in information processing and communication, offering advantages such as microwave-range frequencies, short wavelengths, compact device integration, and notably low energy consumption without Joule heating or charge transport \cite{vogel2007technology, neusser2009magnonics, khitun2010magnonic, lenk2011building, chumak2015magnon, theis2017end}. Despite these benefits, practical implementation of SW-based devices remains challenging due to difficulties in effectively controlling spin wave propagation, which is inherently anisotropic \cite{Pirro2021AdvancesIC, flebus2024roadmap}.
Based on their propagation direction relative to the magnetization orientation, SWs are primarily classified into three categories: backward volume magnetostatic spin wave (BVMSW), magnetostatic surface spin wave (MSSW), and forward volume magnetostatic spin wave (FVMSW) modes \cite{damon1961magnetostatic}. Each mode exhibits distinct propagation characteristics controllable via external magnetic field magnitude and orientation. Specifically, MSSWs and BVMSWs can interconvert in magnetic films subjected to an in-plane magnetic field \cite{Sadovnikov2017spin}, highlighting their suitability as converters or signal processors within complex magnonic circuits. Additional studies \cite{Kasahara2021ferromagnetic} have demonstrated that waveguides with high aspect ratios develop significant demagnetizing fields influenced by shape-induced magnetic anisotropy, causing BVMSWs to become the dominant propagation mode in weak external magnetic fields \cite{bhaskar2020backward}.
Notably, the dispersion relation of BVMSWs presents a unique duality: at low wave numbers, dipole-dipole interactions dominate, resulting in negative group velocities, while at high wave numbers, exchange interactions become prevalent, leading to positive group velocities. Consequently, a single excitation frequency can correspond to two distinct wave vectors, representing different interaction regimes. Clearly distinguishing and characterizing these two spin wave regimes is crucial for mitigating interference and enhancing the reliability of spin logic circuits.

\begin{figure}[htb]
	\centering
	\includegraphics[width=4in]{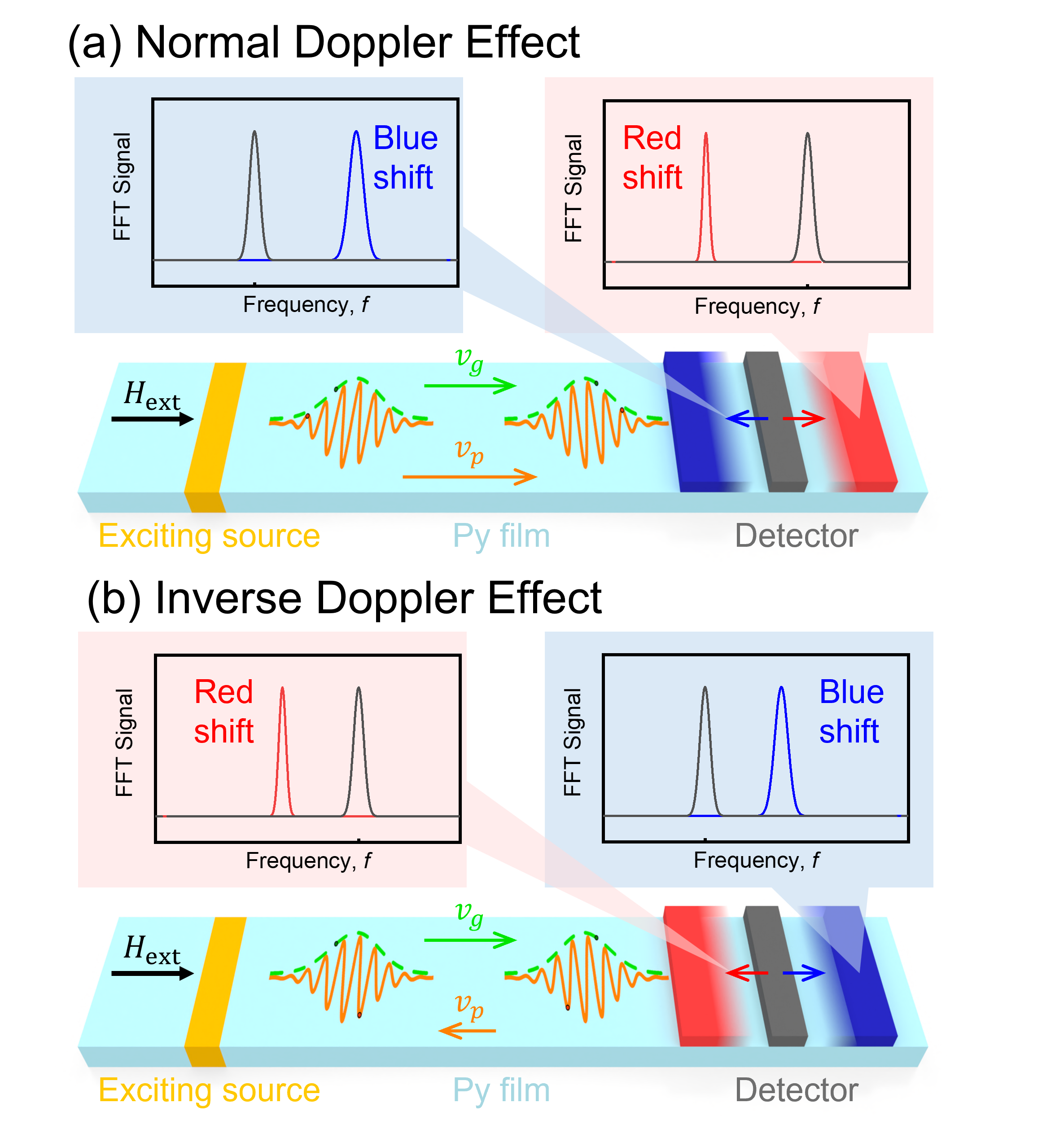}	
	\caption{ The schematic of normal and inverse Doppler effect of BVMSWs in Permalloy film. (a) The normal spin wave Doppler effect with parallel phase velocity ($ v_\mathrm{p} $) and group velocity ($ v_\mathrm{g} $). When the detector approaches the spin wave source, the spin wave will exhibit a blue shift. If the detector moves away from the spin wave source, the spin wave will exhibit a red shift.
    (b) Inverse Doppler effect with anti-parallel $ v_\mathrm{p} $ and $ v_\mathrm{g} $.
    When the detector approaches the spin wave excitation source, the spin wave spectrum will show a red shift. If the detector moves away from the excitation source, the spin wave will show a blue shift.
	}
	\label{figure1}
\end{figure}

Here, we introduce the spin wave Doppler effect, which could effectively control and detect the characteristics of spin waves.\cite{vlaminck2008current,chauleau2014self,xia2016doppler,sugimoto2016observation,song2020backward,kim2021current,nakane2021current,2021Yu,hu2024voltage} In a spin wave system, the frequency of the spin wave will change when the source of the wave and the detector are in relative motion. By adjusting the relative motion between the source of the wave and the detector, the frequency of the spin wave can be changed, thereby achieving the detection of the spin wave. 
Additionally, the blue shift or red shift of spin waves is related to their group velocity and phase velocity. 

The phase and group velocities are
$\boldsymbol{v_p} = \frac{2\pi f}{|\boldsymbol{k}|}\boldsymbol{\hat k}$ and $
\boldsymbol{v_g} = \frac{2\pi \partial f}{\partial{|\boldsymbol{k}|}} \boldsymbol{\hat k}$, where $\boldsymbol{\hat k}$ represents the unit wavevector. % as the same direction as both the phase velocity and the group velocity of the SW. 
For the amplitude-modulated spin wave, when modulation frequency $f_m$ and variation of wave vector $ \boldsymbol{\Delta k} $ are relatively small, the group velocity can be approximated as 
$\boldsymbol{v_g} = \frac{2\pi \partial f}{\partial{|\boldsymbol{k}|}} \boldsymbol{\hat k} \approx \pm \frac{2\pi f_m}{|\boldsymbol{\Delta k}|} \boldsymbol{\hat k}.$
The frequency of the spin wave Doppler effect is  
$f' = \left(1 - \frac{\boldsymbol{v_H} \cdot \boldsymbol{v_p}}{|\boldsymbol{v_p}|^2}\right) f$, where $\boldsymbol{v_H}$ is the velocity of spin wave detector.

As illustrated in Fig.\ref{figure1}(a), when the group velocity and phase velocity of the spin waves are parallel, and the detector approaches the spin wave source ($\boldsymbol{v_H} \cdot \boldsymbol{v_p} < 0$), the spin wave will exhibit a blue shift ($f'>f$). Conversely, if the detector moves away from the spin wave source ($\boldsymbol{v_H} \cdot \boldsymbol{v_p} > 0$), the spin wave will exhibit a red shift ($f'<f$). This is the well-known normal Doppler effect. However, if the group velocity and the phase velocity are anti-parallel, when the detector approaches the spin wave excitation source ($\boldsymbol{v_H} \cdot \boldsymbol{v_p} > 0$), the spin wave spectrum will show a red shift ($f'<f$). When the detector moves away from the excitation source ($\boldsymbol{v_H} \cdot \boldsymbol{v_p} < 0$), the spin wave will show a blue shift ($f'>f$). This unusual phenomenon of spin wave Doppler frequency shift is called the inverse spin wave Doppler effect as shown in Fig.\ref{figure1}(b). 
For BVMSW, spin waves at low wave numbers are expected to exhibit an inverse Doppler effect due to having anti-parallel phase and group velocities\cite{seddon2003observation,stancil2006observation,chumak2010reverse,wessels2016direct}. 
However, these studies typically only consider one wave vector magnon at a single frequency, whereas in reality, two wave vectors should coexist at the same frequency. Previous research on this issue has been scant, and it is rarely mentioned experimentally, mainly because it is difficult to distinguish between spin waves of different wave vectors at the same frequency.
For higher wave numbers, a normal Doppler effect occurs due to the parallel phase and group velocities. This different spin wave Doppler characterization enables more precise differentiation of spin wave vectors with the same frequency.

\begin{figure}[htb]
	\centering
	\includegraphics[width=6in]{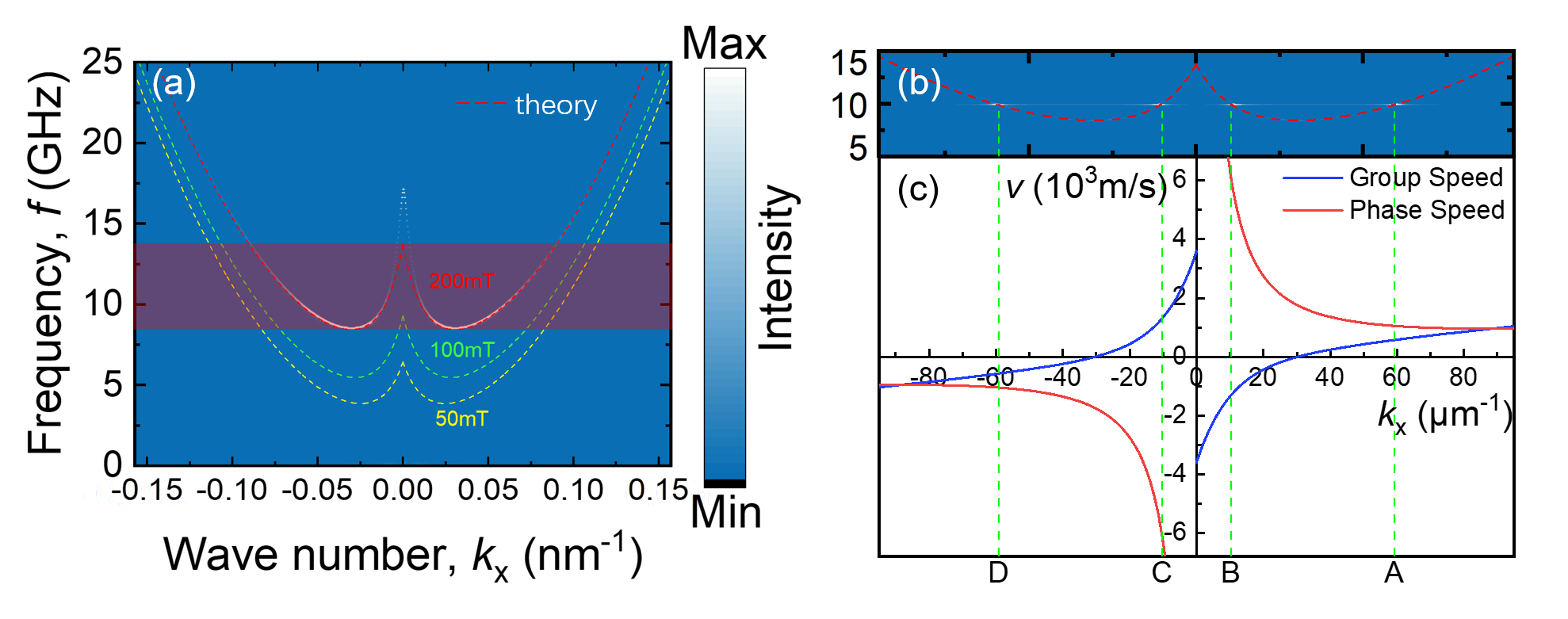}	
	\caption{ (a) The dispersion relation of BVMSW in isotropic Py film by micromagnetic simulation. The dashed lines correspond to the theory calculation with external magnetic field varying from $50\ \mathrm{mT}$ to $200\ \mathrm{mT}$. (b) The dispersion relation of spin waves excited solely by a single frequency of \emph{f}=10 GHz. (c) The theoretically calculated group velocity and phase velocity as a function of wave number. The excited spin waves are highlighted with dashed lines and named A, B, C, and D. 
	}
	\label{figure2}
\end{figure}

To confirm our hypothesis, we perform micromagnetic simulations by MuMax3\cite{vansteenkiste2014design} to study the spin wave Doppler effect of BVMSWs in Permalloy film.
We study a $20000 \times 400 \times 200\ \mathrm{nm}^3$ Permalloy film discretized using $10000 \times 200 \times 1$ finite difference cells. 
The periodic boundary condition is used along \emph{y} axis to avoid the boundary effect. 
The simulation parameters used are as follows: saturation magnetization $M_\mathrm{s} = 0.8 \times 10^6 \ \mathrm{A/m} $, exchange constant $A_\mathrm{ex} = 1.3 \times 10^{-11} \ \mathrm{J/m} $, Gilbert damping $\alpha = 0.006 $.\cite{hu2022significant,Cui2024Magnetic} 
To prevent SW reflection at both ends, $\alpha$ is increased following a squared pattern from 0.006 to 0.1 at both end regions of the Py film ($-10000$ nm \textless \ \emph{x} \textless \ $-8400$ nm, $8400$ nm \textless \ \emph{x} \textless \ $10000$ nm).
An external magnetic field $\mu _0 H_\mathrm{ext} = 200\ \mathrm{mT}$ is applied to magnetize the Py film along +\emph{x} axis. 
Two dimensional Fourier transform on $m_x(x, t)$ in response to a sinc-based excitation field $ \mathrm{h}_0\mathrm{sinc}(2\pi f_\mathrm{c}(t-t_0))\hat{e}_x$ with $\mu _0h_0 = 1\ \mathrm{mT}$, cutoff frequency $f_c = 100\ \mathrm{GHz}$ and $t_0 = 25$ ns, at the center section ($4 \times 400 \times 200\ \mathrm{nm}^3$) of Py film is performed\cite{Kumar2012numerical}. 
At first, the dispersion relations are obtained as shown in Fig.\ref{figure2}(a).  
The dashed lines represent the theoretical calculations from the dispersion relation equations of BVMSWs in isotropic regions given as follow\cite{kalinikos1986theory,wang2022dual,hu2024voltage}: 

\begin{equation*}
    f = F(k) = \frac{\gamma \mu _0}{2\pi}\sqrt{ \left( H_{\mathrm{ext}} + \frac{2A_{\mathrm{ex}}}{\mu_0 M_{\mathrm{s}}}k^2 \right ) \left ( H_{\mathrm{ext}} + \frac{2A_{\mathrm{ex}}}{\mu_0 M_{\mathrm{s}}}k^2 + M_{\mathrm{s}}\frac{1-e^{-kd}}{kd} \right ) }
    \label{f}
\end{equation*} 
where $\gamma$ is the gyromagnetic ratio; $\mu_0$ is the permeability of free space; $k$ is the wave number; $d = 200\ \mathrm{nm}$ is the thickness of the magnetic film.

To enhance our understanding of the dispersion relation, we have plotted the calculated dispersion relations under external magnetic fields of $\mu_0 H_\mathrm{ext} = 50\ \mathrm{mT}$ and $100\ \mathrm{mT}$, as shown in Fig. \ref{figure2}(a).
These calculations reveal that the negative dispersion relations are consistent from the magnetic resonant frequency ($k_x=0$) to the minimum cutoff frequency ($d f/d k_x=0$), at which point the group velocity reaches zero. The cutoff frequency rises with increasing magnetic field strength.
Although micro-magnetic simulations and theoretical calculations generally align across most wave numbers, discrepancies become significant at the smallest wave numbers. 
This mismatch arises partly due to the challenges in simulating long-wavelength spin waves in Permalloy films, which are constrained by their finite size. 
More importantly, while the conditions for de-pinning are met in theory, the simulation can not effectively capture this de-pinning effect, causing the increase in frequencies at the small wave numbers\cite{wang2019spin}.
Here, the simulation parameters of geometry were optimized to minimize deviations at the smallest wave numbers. 
Interestingly,  a single frequency corresponds to two types of spin waves with distinct wave numbers in the range marked as red colour in Fig.\ref{figure2}(a). To verify the unique properties, we present the dispersion relation for spin waves excited solely by a single frequency of $\emph{f}=10 \ \mathrm{GHz}$, as depicted in Fig.\ref{figure2}(b). It is evident that there are four distinct types of magnon vectors.

One of the significant properties of magnons with different wave numbers is the group velocity of spin waves $ \left({\mathrm{d} f}/{\mathrm{d} k}\right) $, the velocity with which the modulation or envelope of the wave propagates through space\cite{bailleul2003propagating}. 
Generally, both the group and phase velocities are the same sign with the wave number, aka Magnon A and D in Fig.\ref{figure2}(c).
However, the group and phase velocities are opposite signs for the magnons B and C. This unique feature mainly comes from the negative group velocity of BVMSWs, where the sign of group velocity is opposite to the wave vector, their propagation direction\cite{stancil2006observation}.

\begin{figure}[htb]
	\centering
	\includegraphics[width=6in]{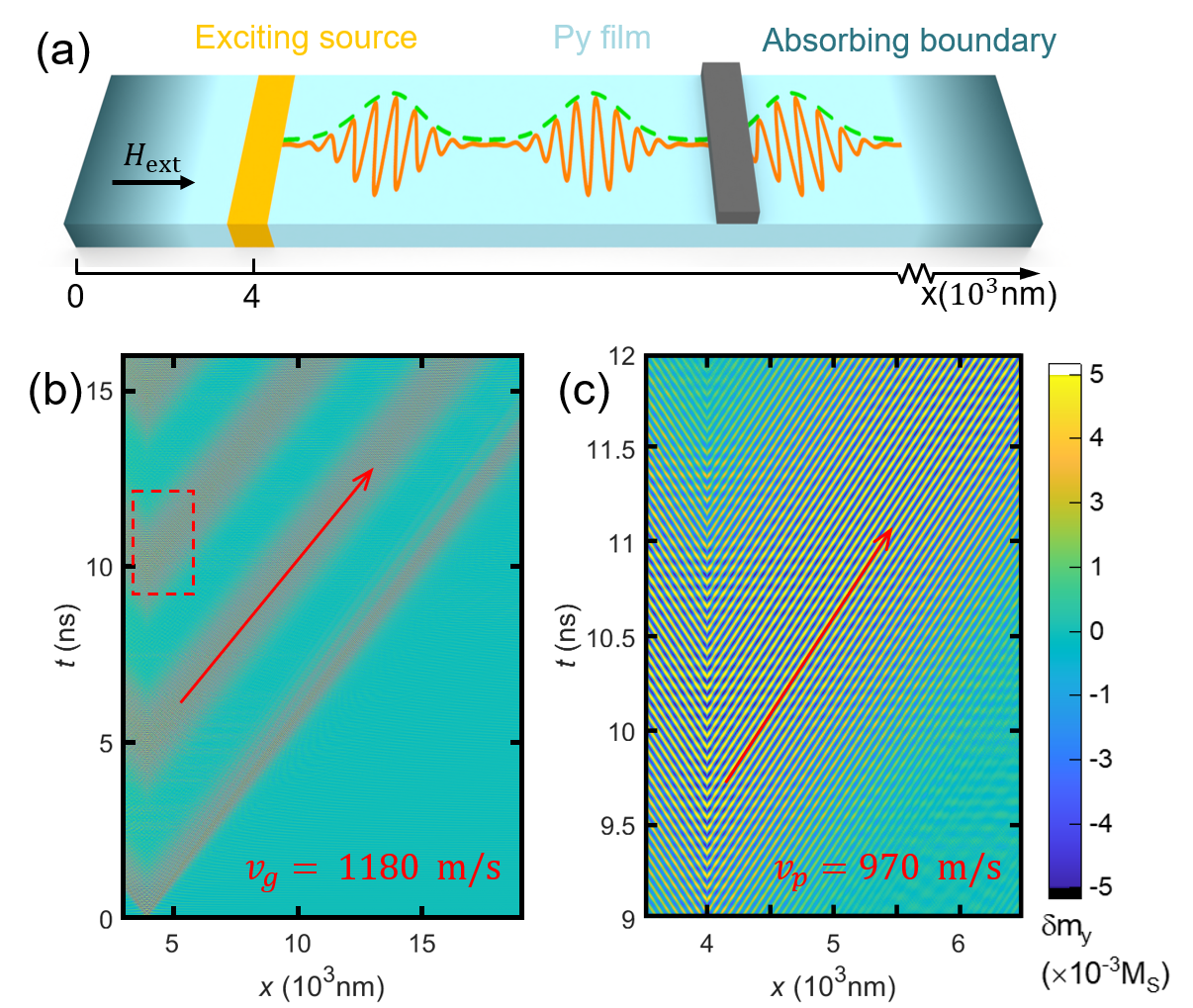}	
	\caption{ (a) The schematic of full-sized structure for simulating the magnetization distribution of time and space.  (b) The magnetization distribution of time and space with the modulated frequency. The speed of the wavepack, identically the group velocity, can be estimated as 1180 m/s. (c) The phase velocity of SW is estimated as 970 m/s from the amplification region. 
	}
	\label{figure3}
\end{figure}

To further study the group and phase velocity properties of spin waves, we set a full-sized $80000 \times 1600 \times 200\ \mathrm{nm}^3$ Py film as shown in Fig.\ref{figure3}(a).  
Additionally, for concentration on SWs propagating along +\emph{x} axis, we move the excitation field along -\emph{x} axis to $x=4\times10^3 \ \mathrm{nm}$ much closer to the absorbing boundary. 
Firstly, only one type of SWs is excited in wavepacks using single-frequency excitation $ h_0\mathrm{sin}(2\pi ft)\times\frac{1}{2}\left ( 1+\mathrm{cos}\left ( 2\pi {f_m t} \right ) \right ) $, where $\mu _0h_0 = 10\ \mathrm{mT}$, $ f = 16\ \mathrm{GHz}$ and $ \mathrm{t}_0 = 5\ \mathrm{ns}$, the modulation frequency $ f_\mathrm{m} = 0.2\ \mathrm{GHz}$.
The image plots of simulated magnetization ($ \delta m_y $) as a function of time and space, as shown in Fig.\ref{figure3}(b,c). 
The group velocity is estimated at $ v_\mathrm{g}=1180\ \mathrm{m/s} $, as observed from the modulated signal propagating through space in Fig. \ref{figure3}(b). The phase velocity is estimated at $ v_\mathrm{p}=970\ \mathrm{m/s} $, based on the propagating carrier wave in Fig. \ref{figure3}(c), originating from the amplification region depicted in Fig. \ref{figure3}(b). These values are consistent with theoretical predictions, clearly confirming the distinguishable difference between the group and phase velocities of spin waves.

\begin{figure}[htb]
	\centering
	\includegraphics[width=6in]{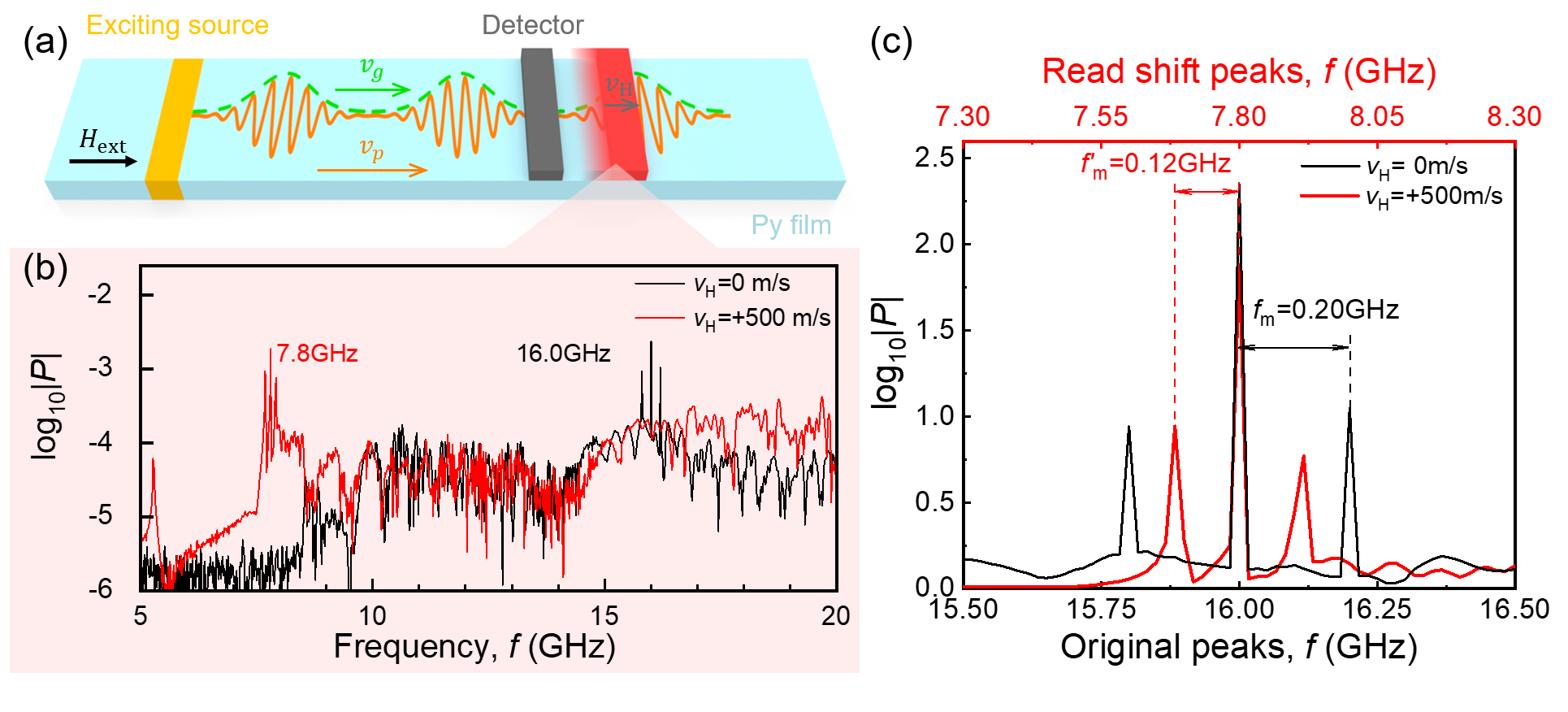}	
	\caption{ (a) The schematic of the simulation for spin wave Doppler effect with the modulated spin waves. (b) The Fourier spectra of signals acquired by fixed detector and detector moving at $\emph{v}_\mathrm{H}= +500\ \mathrm{m/s}$. (c) Enlarged spectra of Fig.(b), illustrating the red shift of secondary peaks. 
	}
	\label{figure4}
\end{figure}

\begin{figure}[htb]
	\centering
	\includegraphics[width=6in]{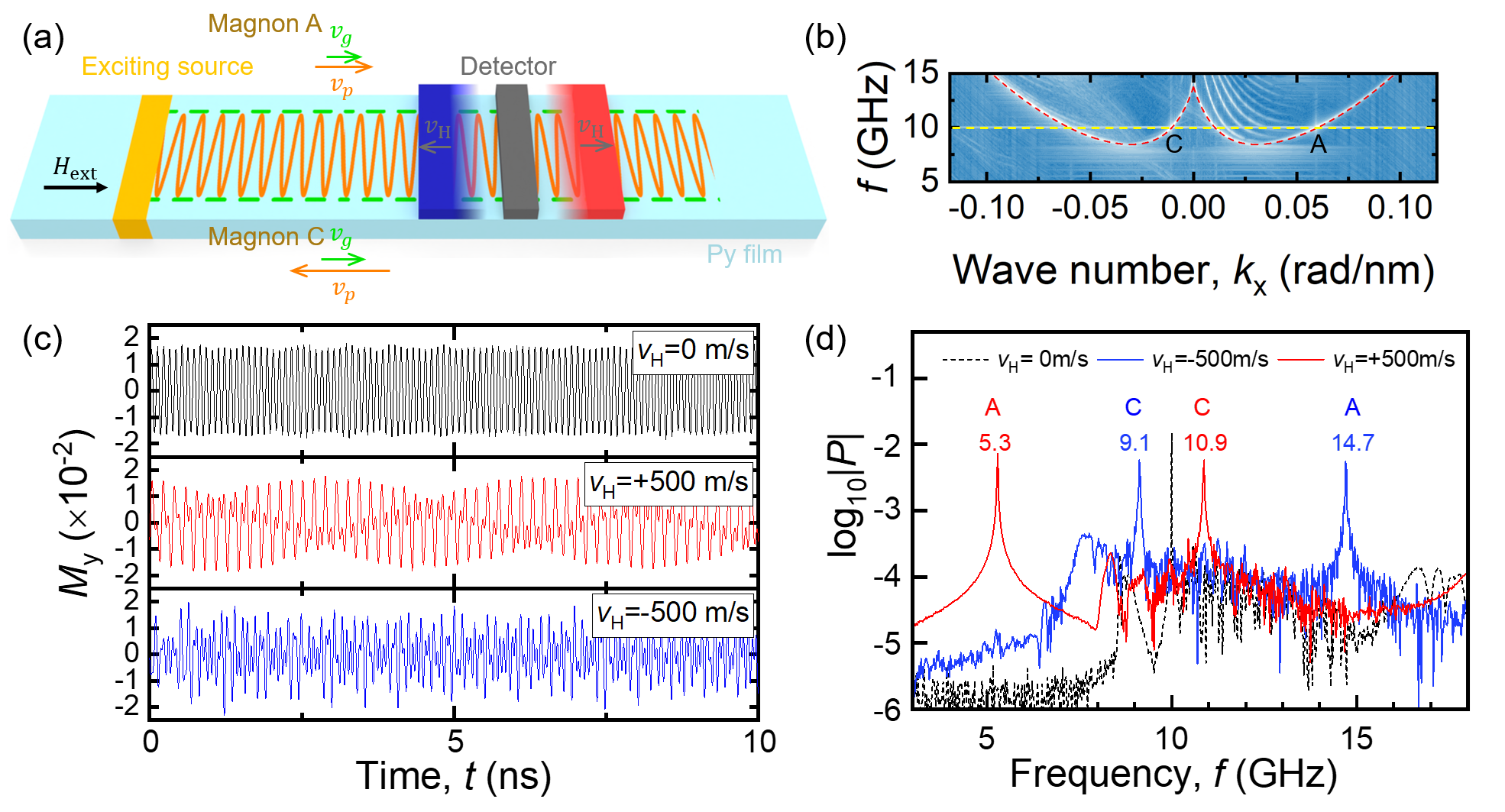}	
	\caption{ (a) The schematic of simulation for a standard \emph{f}=10 GHz SWs. The detectors are stained to distinguish their moving directions based on conventional Doppler Effect. (b) The dispersion relation of spin waves excited solely by a single frequency of \emph{f}=10 GHz at $x=4\times10^3 \ \mathrm{nm}$. Only two types of SWs are excited. (c) Time dependence of magnetization for the various velocities of the detector.  (d) The corresponding Fourier spectra for various velocities of the detector.
	}
	\label{figure5}
\end{figure}

For analysis of the spin wave Doppler effect, we first fixed the detector (grey colour) by sampling from one discretization cell $ \left( \delta x= 2\ nm \right) $, as shown in Fig.\ref{figure4}(a).
Fast Fourier transform on the function $ m_y (t) $ is performed to obtain the spectrum of the unaltered signals of SW. 
And two secondary peaks are observed from the spectrum of the modulated spin waves in Fig.\ref{figure4}(b).  This unique spectrum could be understood by the form function of the modulated spin wave as $ \frac{1}{2}h_0\mathrm{sin}(2\pi ft)+\frac{1}{4}h_0\mathrm{sin}(2\pi (f+f_m)t)+\frac{1}{4}h_0\mathrm{sin}(2\pi (f-f_m)t) $. From Fig.\ref{figure4}(c), we can clearly see that the gap between the main peak and the secondary peak is $ f_m=0.20\ \mathrm{GHz} $, identical to the frequency of modulated frequency of SW. 
Then, we implement detector moving along +\emph{x} axis at velocity $ v_\mathrm{H}=+500\ \mathrm{m/s} $ by continuous movement of sampling point on the Py film over one discretization cell during each time window $ \left ( \delta t= \delta x/v_\mathrm{H} \right ) $ in the simulation\cite{hu2024voltage}. 
The spin wave spectrum is obtained for moving detector by performing an FFT on $m_y(t)$, as illustrated by the red curve in Fig.\ref{figure4}(b). 
The main peak ($f'$) is read as $ 7.8\ \mathrm{GHz} $, a significant red shift of SW. The phase velocity could be evaluated as the value of $ v_\mathrm{p}=970\ \pm 5\ \mathrm{m/s} $ based on the Doppler frequency shift formula $f'=\left(1 - \frac{\boldsymbol{v_H} \cdot \boldsymbol{v_p}}{|\boldsymbol{v_p}|^2}\right) f$. 
The change of gap between the main peak and the secondary peak is obtained as $f_m' = 0.12\ \mathrm{GHz} $ by the enlarged spectra with aligned main peaks in Fig.\ref{figure4}(c). 
Actually, the frequency shift of the secondary peak gap can be used to estimate the group velocity of spin waves. 
The obtained group velocity $ v_\mathrm{g}=1250\ \pm \ 80\ 
 \mathrm{m/s} $ from  the formula $f_m'=\left(1 - \frac{\boldsymbol{v_H} \cdot \boldsymbol{v_g}}{|\boldsymbol{v_g}|^2}\right) f_m$  with the value of  $ f_m'=0.12\ \mathrm{GHz} $ is nearly consistent with the value obtained in Fig.\ref{figure3}. This error arises from the fact that the frequency resolution of the FFT spectrum is only 0.01 GHz, and the frequency gap $ f_m $ cannot be regarded as an infinitesimal quantity as shown in Fig.\ref{figure4}(d). The result indicates that the group velocity-induced spin wave Doppler effect could be precisely analysed using the frequency shift difference of the main and secondary peaks, while also presents a straightforward method for approximating the group velocity of SWs.

\begin{figure}[htb]
	\centering
	\includegraphics[width=5in]{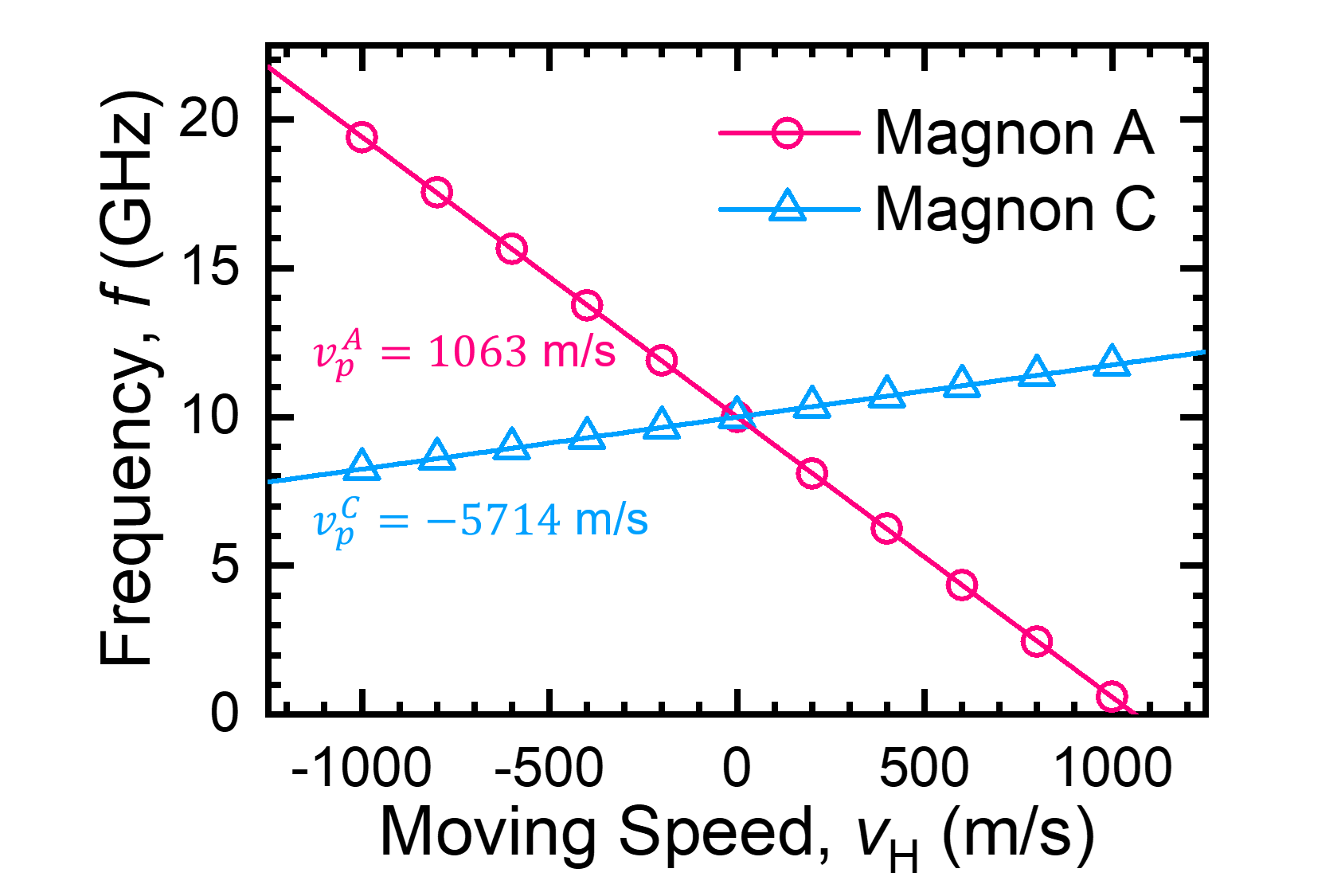}	
	\caption{ The simulated frequency shifts across various detector velocities for magnon A and C. The solid lines correspond to the theoretical calculation. 
	}
	\label{figure6}
\end{figure}

A single-frequency excitation source, defined by $ h_0\sin(2\pi ft) $ where $\mu _0h_0 = 10\ \mathrm{mT}$ and $ f = 10\ \mathrm{GHz}$, was utilized in subsequent simulations. This frequency excites two types of SWs, as depicted by the dispersion relation of backward-volume magnetostatic spin waves in Fig.\ref{figure2}(a). The dispersion relation of BVMSW shows two highlighted spots, which correspond to the magnon models A and C in Fig.\ref{figure5}(b). In contrast, magnon modes B and D are suppressed due to spatial limitations along the negative x-axis. The design of the geometry thus plays a crucial role in accurately analyzing the spin wave Doppler effect.
Initially, the time-dependent magnetization spectra were recorded by the detector under various motion scenarios, as illustrated in Fig.\ref{figure5}(c). For a stationary detector ($\emph{v}_{\mathrm{H}}= 0\ \mathrm{m/s}$), the $\delta m_y (t)$ curve exhibits a standard sine function pattern, aligning with the spin wave's excitation function. The frequency spectrum for this case shows a single peak at 10 GHz (black curve in Fig.\ref{figure5}(d). Conversely, with the detector moving at $\emph{v}_{\mathrm{H}}= 500\ \mathrm{m/s}$, the $\delta m_y (t)$ curve resembles a modulated wave, as shown in Fig.\ref{figure5}(c). This motion results in two distinct frequency peaks at 5.3 GHz and 10.9 GHz, depicted by the red curve in Fig.\ref{figure5}(d).
The red-shifted frequency peak at $5.3\ \mathrm{GHz}$ can be attributed to the normal Doppler effect, where the detector is positioned far from the source, primarily influenced by magnon A. Conversely, magnon C, moving in a scenario where phase and group velocities are anti-parallel, contributes to a blue-shifted frequency peak at $10.9\ \mathrm{GHz}$, indicative of the inverse Doppler effect. Reversing the detector's movement along the -\emph{x} axis to a velocity of $v_\mathrm{H}=+500\ \mathrm{m/s}$ also distinctly produces two peaks in the frequency spectrum (blue curve in Fig.\ref{figure5}(d)). A prominent blue-shifted peak at $14.7\ \mathrm{GHz}$ due to the normal Doppler effect from magnon A, and a smaller red-shifted peak from magnon C, consistent with the inverse Doppler effect.
To accurately determine the phase velocities of magnons A and C, we simulated the frequency shifts across various detector velocities as displayed in Fig.\ref{figure6}. By employing linear fitting, we calculated the phase velocities of magnon A and C are $1063\ \mathrm{m/s}$ and $-5714\ \mathrm{m/s}$, respectively. The results align well with those predicted from the dispersion relation in Fig.\ref{figure2}. The negative phase velocity value for magnon C confirms the inverse Doppler effect. These observations offer a robust method for differentiating between the two magnon models through their Doppler shifts.

In summary, our study has revealed, in the case of BVMSWs, spin waves at low wave numbers display an inverse Doppler effect because their phase and group velocities are anti-parallel. Conversely, at high wave numbers, a normal Doppler effect occurs due to the parallel alignment of phase and group velocities. Analyzing the spin wave Doppler effect offers a novel perspective for understanding intrinsic interactions and can also help mitigate serious interference issues in the design of spin logic circuits.

\begin{acknowledgments}
This work is partially supported by Shenzhen Science and Technology Program \\(JCYJ20240813113228037).
\end{acknowledgments}

%\newpage

\bibliographystyle{unsrt}
\bibliography{Bib}

\begin{thebibliography}{10}

\bibitem{Bloch1932zur}
Felix Bloch.
\newblock {\em Zur Theorie des Austauschproblems und der Remanenzerscheinung der Ferromagnetika}, pages 295--335.
\newblock Springer Berlin Heidelberg, Berlin, Heidelberg, 1932.

\bibitem{Holstein1940field}
T.~Holstein and H.~Primakoff.
\newblock Field dependence of the intrinsic domain magnetization of a ferromagnet.
\newblock {\em Phys. Rev.}, 58:1098--1113, 1940.

\bibitem{Dyson1956general}
Freeman~J. Dyson.
\newblock General theory of spin-wave interactions.
\newblock {\em Phys. Rev.}, 102:1217--1230, 1956.

\bibitem{vogel2007technology}
Eric Vogel.
\newblock Technology and metrology of new electronic materials and devices.
\newblock {\em Nature nanotechnology}, 2(1):25--32, 2007.

\bibitem{neusser2009magnonics}
Sebastian Neusser and Dirk Grundler.
\newblock Magnonics: Spin waves on the nanoscale.
\newblock {\em Advanced materials}, 21(28):2927--2932, 2009.

\bibitem{khitun2010magnonic}
Alexander Khitun, Mingqiang Bao, and Kang~L Wang.
\newblock Magnonic logic circuits.
\newblock {\em Journal of Physics D: Applied Physics}, 43(26):264005, 2010.

\bibitem{lenk2011building}
Benjamin Lenk, Henning Ulrichs, Fabian Garbs, and Markus M{\"u}nzenberg.
\newblock The building blocks of magnonics.
\newblock {\em Physics Reports}, 507(4-5):107--136, 2011.

\bibitem{chumak2015magnon}
Andrii~V Chumak, Vitaliy~I Vasyuchka, Alexander~A Serga, and Burkard Hillebrands.
\newblock Magnon spintronics.
\newblock {\em Nature physics}, 11(6):453--461, 2015.

\bibitem{theis2017end}
Thomas~N Theis and H-S~Philip Wong.
\newblock The end of moore's law: A new beginning for information technology.
\newblock {\em Computing in science \& engineering}, 19(2):41--50, 2017.

\bibitem{Pirro2021AdvancesIC}
Philipp Pirro, Vitaliy~I. Vasyuchka, Alexander~A. Serga, and Burkard Hillebrands.
\newblock Advances in coherent magnonics.
\newblock {\em Nature Reviews Materials}, 6:1114 -- 1135, 2021.

\bibitem{flebus2024roadmap}
Benedetta Flebus, Dirk Grundler, Bivas Rana, Yoshichika Otani, Igor Barsukov, Anjan Barman, Gianluca Gubbiotti, Pedro Landeros, Johan Akerman, Ursula~S Ebels, Philipp Pirro, V~E Demidov, Katrin Schultheiss, Gyorgy Csaba, Qi~Wang, Dmitri~E. Nikonov, Florin Ciubotaru, Ping Che, Riccardo hertel, Teruo Ono, Dmytro Afanasiev, Johan~H Mentink, Theo Rasing, Burkard Hillebrands, Silvia Viola~Kusminskiy, Wei Zhang, Chunhui~Rita Du, Aurore Finco, Toeno van~der Sar, Yunqiu~Kelly Luo, Yoichi Shiota, Joseph Sklenar, Tao Yu, and Jinwei Rao.
\newblock The 2024 magnonics roadmap.
\newblock {\em Journal of Physics: Condensed Matter}, 2024.

\bibitem{damon1961magnetostatic}
Richard~W Damon and JR~Eshbach.
\newblock Magnetostatic modes of a ferromagnet slab.
\newblock {\em Journal of Physics and Chemistry of Solids}, 19(3-4):308--320, 1961.

\bibitem{Sadovnikov2017spin}
A.~V. Sadovnikov, C.~S. Davies, V.~V. Kruglyak, D.~V. Romanenko, S.~V. Grishin, E.~N. Beginin, Y.~P. Sharaevskii, and S.~A. Nikitov.
\newblock Spin wave propagation in a uniformly biased curved magnonic waveguide.
\newblock {\em Phys. Rev. B}, 96:060401, 2017.

\bibitem{Kasahara2021ferromagnetic}
Kenji Kasahara, Ryusei Akamatsu, and Takashi Manago.
\newblock {Ferromagnetic-waveguide width dependence of propagation properties for magnetostatic surface spin waves}.
\newblock {\em AIP Advances}, 11(4):045308, 2021.

\bibitem{bhaskar2020backward}
UK~Bhaskar, Giacomo Talmelli, Florin Ciubotaru, Christoph Adelmann, and Thibaut Devolder.
\newblock Backward volume vs damon--eshbach: A traveling spin wave spectroscopy comparison.
\newblock {\em Journal of Applied Physics}, 127(3), 2020.

\bibitem{vlaminck2008current}
Vincent Vlaminck and Matthieu Bailleul.
\newblock Current-induced spin-wave doppler shift.
\newblock {\em Science}, 322(5900):410--413, 2008.

\bibitem{chauleau2014self}
J-Y Chauleau, HG~Bauer, HS~K{\"o}rner, J~Stigloher, M~H{\"a}rtinger, G~Woltersdorf, and CH~Back.
\newblock Self-consistent determination of the key spin-transfer torque parameters from spin-wave doppler experiments.
\newblock {\em Physical Review B}, 89(2):020403, 2014.

\bibitem{xia2016doppler}
Hong Xia, Jie Chen, Xiaoyan Zeng, and Ming Yan.
\newblock Doppler effect in a solid medium: Spin wave emission by a precessing domain wall drifting in spin current.
\newblock {\em Physical Review B}, 93(14):140410, 2016.

\bibitem{sugimoto2016observation}
Satoshi Sugimoto, Mark~C Rosamond, Edmund~H Linfield, and Christopher~H Marrows.
\newblock {Observation of spin-wave Doppler shift in $\mathrm{Co_{90}Fe_{10}/Ru}$ micro-strips for evaluating spin polarization}.
\newblock {\em Applied Physics Letters}, 109(11), 2016.

\bibitem{song2020backward}
Wenjie Song, Xiansi Wang, Wenfeng Wang, Changjun Jiang, Xiangrong Wang, and Guozhi Chai.
\newblock {Backward magnetostatic surface spin waves in coupled $\mathrm{Co/FeNi}$ bilayers}.
\newblock {\em Physica Status Solidi (RRL)--Rapid Research Letters}, 14(8):2000118, 2020.

\bibitem{kim2021current}
Dong-Hyun Kim, Se-Hyeok Oh, Dong-Kyu Lee, Se~Kwon Kim, and Kyung-Jin Lee.
\newblock Current-induced spin-wave doppler shift and attenuation in compensated ferrimagnets.
\newblock {\em Physical Review B}, 103(1):014433, 2021.

\bibitem{nakane2021current}
Jotaro~J Nakane and Hiroshi Kohno.
\newblock Current-induced spin-wave doppler shift in antiferromagnets.
\newblock {\em Journal of the Physical Society of Japan}, 90(10):103705, 2021.

\bibitem{2021Yu}
Tao Yu, Chen Wang, Michael~A. Sentef, and Gerrit E.~W. Bauer.
\newblock {Spin-Wave Doppler Shift by Magnon Drag in Magnetic Insulators}.
\newblock {\em Phys. Rev. Lett.}, 126:137202, 2021.

\bibitem{hu2024voltage}
Shaojie Hu, Kang Wang, Tai Min, and Takashi Kimura.
\newblock Voltage-controlled spin-wave doppler shift in a ferromagnetic/ferroelectric heterojunction.
\newblock {\em Physical Review Applied}, 22(1):014085, 2024.

\bibitem{seddon2003observation}
N~Seddon and T~Bearpark.
\newblock Observation of the inverse doppler effect.
\newblock {\em Science}, 302(5650):1537--1540, 2003.

\bibitem{stancil2006observation}
Daniel~D Stancil, Benjamin~E Henty, Ahmet~G Cepni, and JP~Van’t~Hof.
\newblock Observation of an inverse doppler shift from left-handed dipolar spin waves.
\newblock {\em Physical Review B—Condensed Matter and Materials Physics}, 74(6):060404, 2006.

\bibitem{chumak2010reverse}
AV~Chumak, P~Dhagat, A~Jander, AA~Serga, and B~Hillebrands.
\newblock Reverse doppler effect of magnons with negative group velocity scattered from a moving bragg grating.
\newblock {\em Physical Review B—Condensed Matter and Materials Physics}, 81(14):140404, 2010.

\bibitem{wessels2016direct}
Philipp Wessels, Andreas Vogel, Jan-Niklas T{\"o}dt, Marek Wieland, Guido Meier, and Markus Drescher.
\newblock Direct observation of isolated damon-eshbach and backward volume spin-wave packets in ferromagnetic microstripes.
\newblock {\em Scientific reports}, 6(1):22117, 2016.

\bibitem{vansteenkiste2014design}
Arne Vansteenkiste, Jonathan Leliaert, Mykola Dvornik, Mathias Helsen, Felipe Garcia-Sanchez, and Bartel Van~Waeyenberge.
\newblock The design and verification of mumax3.
\newblock {\em AIP advances}, 4(10), 2014.

\bibitem{hu2022significant}
Shaojie Hu, Xiaomin Cui, Kang Wang, Satoshi Yakata, and Takashi Kimura.
\newblock Significant modulation of vortex resonance spectra in a square-shape ferromagnetic dot.
\newblock {\em Nanomaterials}, 12(13), 2022.

\bibitem{Cui2024Magnetic}
Xiaomin Cui, Shaojie Hu, Yohei Hidaka, Satoshi Yakata, and Takashi Kimura.
\newblock Magnetic vortex polarity reversal induced gyrotropic motion spectrum splitting in a ferromagnetic disk.
\newblock {\em Journal of Physics D: Applied Physics}, 57(39):395002, 2024.

\bibitem{Kumar2012numerical}
Dheeraj Kumar, Oleksandr Dmytriiev, Sabareesan Ponraj, and Anjan Barman.
\newblock Numerical calculation of spin wave dispersions in magnetic nanostructures.
\newblock {\em Journal of Physics D: Applied Physics}, 45(1):015001, 2011.

\bibitem{kalinikos1986theory}
BA~Kalinikos and AN~Slavin.
\newblock Theory of dipole-exchange spin wave spectrum for ferromagnetic films with mixed exchange boundary conditions.
\newblock {\em Journal of Physics C: Solid State Physics}, 19(35):7013, 1986.

\bibitem{wang2022dual}
Kang Wang, Shaojie Hu, Fupeng Gao, Miaoxin Wang, and Dawei Wang.
\newblock Dual function spin-wave logic gates based on electric field control magnetic anisotropy boundary.
\newblock {\em Applied Physics Letters}, 120(14), 2022.

\bibitem{wang2019spin}
Q~Wang, B~Heinz, R~Verba, M~Kewenig, P~Pirro, M~Schneider, T~Meyer, B~L{\"a}gel, C~Dubs, T~Br{\"a}cher, et~al.
\newblock Spin pinning and spin-wave dispersion in nanoscopic ferromagnetic waveguides.
\newblock {\em Physical review letters}, 122(24):247202, 2019.

\bibitem{bailleul2003propagating}
Matthieu Bailleul, Dominik Olligs, and Claude Fermon.
\newblock Propagating spin wave spectroscopy in a permalloy film: A quantitative analysis.
\newblock {\em Applied Physics Letters}, 83(5):972--974, 2003.

\end{thebibliography}

\end{document}